\documentclass[10pt, aps,prc,twocolumn,superscriptaddress,preprintnumbers,
amsmath, 
floatfix,
longbibliography,
nofootinbib
]{revtex4-1}
\usepackage[T1]{fontenc}
\usepackage[utf8x]{inputenc} 
\usepackage{adjustbox}          
\usepackage[caption=false]{subfig}
\usepackage{url}
\usepackage{color, xcolor}
\usepackage{float}
\usepackage[pdftex,colorlinks=true, linkcolor = blue, citecolor=blue,urlcolor=blue, bookmarksnumbered=true, bookmarksopen=true]{hyperref}
\usepackage{longtable}
\usepackage{amsfonts}
\usepackage{dsfont}
\usepackage{wrapfig,bm} 
\usepackage[normalem]{ulem}
\usepackage{MnSymbol}
\usepackage{cancel}
\usepackage[normalem]{ulem}

\newcommand{\beq}{\begin{equation}}
\newcommand{\eeq}{\end{equation}}
\newcommand{\bea}{\begin{eqnarray}}
\newcommand{\eea}{\end{eqnarray}}

\begin{document}
\title{Pair Transfer and Reaction Dynamics in $^{40,48}$Ca + $^{96}$Zr Collisions Below the Coulomb Barrier}
  
\author{Ibrahim Abdurrahman}
  \affiliation{ Facility for Rare Isotope Beams, Michigan State University, East Lansing, Michigan 48824, USA}
 \affiliation{ Theoretical Division, Los Alamos National Laboratory, Los Alamos, New Mexico 87545, USA}

 \author{Andrzej Makowski}
 \affiliation{Faculty of Physics, Warsaw University of Technology, \\
ulica Koszykowa 75, Warsaw, 00-662, Poland}

 \author{Guillaume Scamps}
 \affiliation{ Université de Toulouse, CNRS/IN2P3, L2IT, Toulouse, France }

 \author{Kyle Godbey}
 \affiliation{ Facility for Rare Isotope Beams, Michigan State University, East Lansing, Michigan 48824, USA}

  \author{Piotr Magierski}
 \affiliation{Faculty of Physics, Warsaw University of Technology, \\
ulica Koszykowa 75, Warsaw, 00-662, Poland}
 \affiliation{Physics Department, University of Washington, \\ 3910 15th Ave. NE, Seattle, WA 98195-1560, USA}
 
\date{\today}

\begin{abstract}


Sub-barrier fusion reactions are ideal for probing the effects of pairing correlations on simultaneous neutron transfer. Previous calculations using the BCS approximation showed an enhancement of pair transfer, relative to treatments with no pairing, but failed to reproduce the observed enhancement factor between one- and two-neutron transfer probabilities. This work aims to microscopically investigate the dynamics of $^{40,48}$Ca + $^{96}$Zr head-on collisions below the Coulomb barrier, focusing on the role of pairing correlations in neutron transfer. We employ time-dependent energy density functional theory extended to superfluid systems, TDSLDA. Transfer probabilities, including contributions to specific $K$-angular momentum projections, are extracted using projection operators and compared to results from calculations without pairing. Our calculations show that pairing is correlated to the dynamic deformability of the nucleus, which influences mean neutron transfer in sub-barrier reactions. We also show that TDSLDA reproduces the experimentally observed enhancement factor by significantly increasing the probability of transferring a neutron pair in the $K = 0$ spin channel. These results confirm the strong influence of pairing and structure on sub-barrier multi-nucleon transfer, and demonstrate that TDSLDA provides a reliable microscopic framework for describing the interplay between nuclear superfluidity and reaction dynamics.


\end{abstract}

\preprint{LA-UR-25-31452}

\maketitle


 \vspace{0.5cm}

Transfer reactions at energies below and near the Coulomb barrier provide a sensitive probe of nuclear structure and correlations, particularly the role of pairing. At such low energies, the interacting nuclei barely touch, and nucleon transfer proceeds predominantly through quantum tunneling and typically involves only a few particles. 
Pairing correlations are expected to enhance the probability of two-nucleon transfer, which in their absence would occur mainly as a sequential process. If pairing correlations generate the long-range order associated with the superfluid phase, then two-nucleon transfer can be viewed as a manifestation of the collective flow of Cooper pairs between finite superconducting systems. Various theoretical studies have explored this analogy,
predicting both DC~\cite{josephson,DIETRICH1970428,DIETRICH1971480,DIETRICH1971201} and AC Josephson currents~\cite{Potel:2021,PhysRevC.105.L061602}. However, clear and unambiguous experimental evidence for these effects remains elusive.

Experimentally, it has been observed that pairing enhances the two-neutron transfer cross section in reactions such as $^{40}\text{Ca} + {}^{96}\text{Zr}$~\cite{Corradi:2011} and $^{60}\text{Ni} + {}^{116}\text{Sn}$~\cite{Montanari:2014}. A key observable to quantify this effect is the enhancement ratio $P_{2n} / P_{1n}^2$, where $P_{1n}$ and $P_{2n}$ denote the one- and two-neutron transfer probabilities, respectively. In the absence of pairing correlations, this ratio is theoretically expected to lie between 0.25 and 0.5~\cite{Scamps:2013,Hagino:2015}, whereas experimental values are found to be significantly larger—about $3$ for $^{40}\text{Ca} + {}^{96}\text{Zr}$~\cite{Corradi:2011} and 5.5 for $^{60}\text{Ni} + {}^{116}\text{Sn}$~\cite{Montanari:2014}.

In this study, time-dependent density functional theory (TDDFT) extended to superfluid systems, or more specifically, the time-dependent superfluid local density approximation (TDSLDA), see Refs.~\cite{PhysRevC.65.051305,bulgac2012,annurev-nucl-102212-170631,magierski2019nuclear,Shi:2021,Magierski:2025} for more details, has been used to investigate head-on collisions of $^{40,48}$Ca + $^{96}$Zr in the sub-barrier regime. TDSLDA has previously been applied successfully to describe 
low-energy nuclear collisions~\cite{Magierski:2017,Sekizawa:2017,Sekizawa:2017b,Barton:2020,Magierski:2022,Makowski:2023}, 
and nuclear  fission~\cite{Bulgac:2016,Bulgac:2019c,Bulgac:2019d,Bulgac:2020,Bulgac:2020b,Bulgac:2021,Bulgac:2022e,Bulgac:2024,Abdurrahman:2024,Bulgac:2025}, 
as well as dynamics of superfluid neutron matter in the neutron star crust and nonequilibrium
superfluid phenomena in ultracold atomic gases
(see Refs.\cite{Magierski2024,Magierski:2025} and references therein). For the first time, the framework is used to investigate one and two-neutron transfer probabilities in low-energy heavy-ion collisions.  


These quantities have been investigated within other TDDFT frameworks such as TDHF without pairing correlations~\cite{Simenel:2010,Sekizawa:2013,Wu:2019,Godbey2020}, TDHF + the BCS approximation~\cite{Scamps:2013}, and covariant density functional theory~\cite{Zhang:2024,Li:2024}, which sometimes also includes pairing correlations via the BCS approximation. The TDHF+BCS approximation, or equivalently, CbTDHFB~\cite{ebata2010} has proven to be a valuable tool for incorporating pairing correlations in nuclear dynamics~\cite{scamps2018,QIANG2025139248}. Its low computational cost, relative to TDSLDA, and success in describing the evolution of the one-body density make it especially useful for exploring the large parameter space available in nuclear collisions: colliding nuclei, collision energies, orientations, and impact parameters. 
TDSLDA, on the other hand, provides full access to the dynamics of the pairing field, in contrast to the 
TDHF+BCS/CbTDHFB frameworks, where pairing is treated in a simplified manner and leads to violations of the one-body continuity equation~\cite{scamps2012}.
This difference has tangible consequences demonstrated in studies of nuclear collisions~\cite{PhysRevC.94.014610,Magierski:2017,Magierski:2022} and induced fission~\cite{Bulgac:2019c,scamps2015superfluid,Ren:2022}.


The primary motivation of this study is to examine the role of pairing correlations (TDSLDA) on central collisions of $^{40,48}$Ca + $^{96}$Zr below the fusion threshold.  As such, we also simulated collisions of the same nuclei without pairing (TDHF) as reference states, using the same code. We used the nuclear energy density functional (NEDF) SeaLL1~\cite{Bulgac:2018}.  The choice of the NEDF will certainly affect quantitative results, hence we will focus more on trends and qualitative features below. Further technical details are provided in the Supplemental materials~\cite{Supplement:2026}, which includes Refs.~\cite{Corradi:2011,Scamps:2013,Ring:2004,ZHANG:2025,Bertsch12,Robledo:2025}.

\begin{figure}[h] \includegraphics[width= \columnwidth]{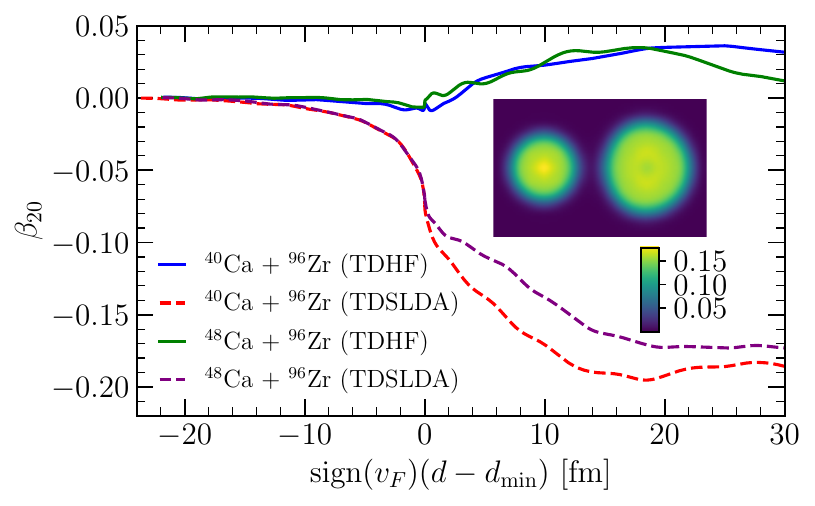}  \caption{ \label{fig:deform} The deformation parameter $\beta_{20}$ of $^{96}$Zr is shown as a function of the separation distance $d = \bm{r}_{cm,R} - \bm{r}_{cm,L}$ between the two nuclei minus the minimum separation distance reached during the collision $d_{\mathrm{min}}$. The separation distance is multiplied by sign($v_F$), which is -1 when the nuclei are moving towards each other, and +1 when they are moving away from each other. The inset shows the density slices in fm$^{-3}$ of a TDSLDA simulation of $^{40}$Ca + $^{96}$Zr at the minimum separation distance. All collisions are shown at $E_{\mathrm{cm}}$ = 95 MeV.  
 } \end{figure}

The first noticeable effect of pairing correlations is their impact on the dynamically induced quadrupole deformation of $^{96}$Zr in collisions with both Ca isotopes, see the top panel of Fig.~\ref{fig:deform}.  
In the TDSLDA framework, $^{96}$Zr gradually develops an oblate quadrupole moment at large separation distances, induced by the long-range Coulomb field generated by the calcium nuclei, which modifies the proton Fermi surface. The magnitude of this moment continues to increase for a short time after the nuclei make contact, eventually stabilizing at a value of approximately $\approx-0.15$ to $-0.2$. 
This behavior reflects the enhanced susceptibility of the system to oblate deformation. Indeed, various mean-field calculations
\cite{PhysRevC.100.044315,PhysRevC.94.044314,SKALSKI1997282,PhysRevC.96.064303,ntx3-nyf8}
predict competing oblate and prolate minima in zirconium isotopes around $N\approx 56$. A calculation involving the SeaLL1 EDF has also been included in the Supplemental materials~\cite{Supplement:2026}. 
An increase in softness toward triaxial deformation has also been reported in the shell model~\cite{Kremer:2016}. 
Note that in the current study, axial symmetry is preserved during evolution, and, as a result, the triaxial degree of freedom remains unexplored. 

\begin{figure}[h] \includegraphics[width= \columnwidth]{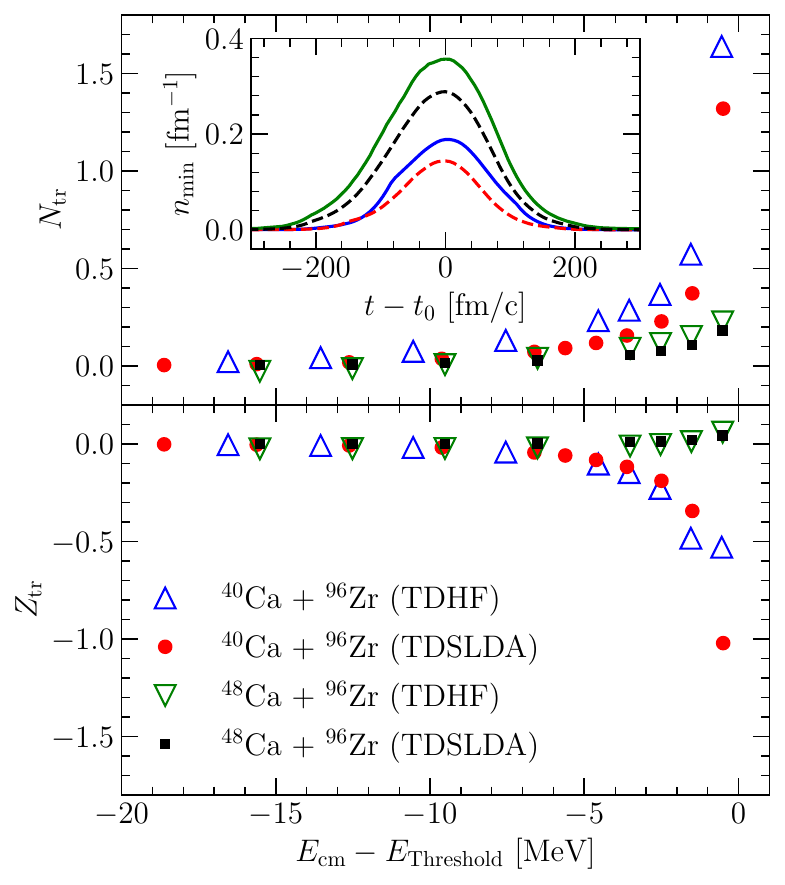}  \caption{ \label{fig:ntrans}  (Top/bottom panel) The mean transfer of (neutrons/protons) to the Ca isotopes is shown as a function of the collision energy with respect to the threshold energy provided in Table~\ref{tbl:ethres}. At the same (unshifted) collision energy, the mean neutron transfer is surprisingly similar for both Ca isotopes. This is likely the result of two competing trends; equilibration of the N/Z ratio, which enhances transfer for the $^{40}$Ca + $^{96}$Zr reaction, while the larger neutron skin of $^{48}$Ca, relative to $^{40}$Ca, enhances transfer for the $^{48}$Ca + $^{96}$Zr reaction. This is further supported by the average proton transfer, which is transferred from $^{40}$Ca to $^{96}$Zr. In this case, there are significantly more protons transferred than for the $^{48}$Ca + $^{96}$Zr reaction at the same (unshifted) collision energy. The inset shows $n_{\mathrm{min}}$, defined by Eq.~\ref{eqn:nmin}, as a function of $t$ with respect to $t_0$, which is defined at the time when $n_{\mathrm{min}}$ reaches its maximum value. The results in the inset correspond to collisions at an energy $E_{\mathrm{cm}}$ = 95 MeV, and the colors correspond to the same collisions labeled in the legend in the bottom panel}  \end{figure}

In TDHF, $^{96}$Zr remains spherical until it reaches the minimum separation distance, where the tails of nuclear densities have the greatest overlap, after which the nuclei separate and it begins to develop a prolate quadrupole moment along the collision axis. Unlike in TDSLDA, in TDHF the wavefunction of $^{96}$Zr is an ordinary Slater determinant, in which the lowest orbitals are fully occupied. As a result, $^{96}$Zr is much more resilient to changing it's ground-state deformation, until the nuclei make contact and can exchange nucleons. At a fixed separation distance, the prolate deformation allows for a greater overlap of the nuclear densities, relative to the oblate deformation, extending the time for the nuclei to interact.  This is shown in the inset of Fig.~\ref{fig:ntrans} where the minimum integrated neutron density between the nuclei is given by,
\begin{equation} \label{eqn:nmin}
    n_\mathrm{min} = \mathrm{min} \Big ( \int n_n (x,y,z) dx dy \Big ),
\end{equation} 
with $z$ representing the collision axis. Both the greater overlap of the densities and the longer interaction time lead to a larger mean transfer for $^{40,48}$Ca + $^{96}$Zr in TDHF relative to TDSLDA, as shown in Fig.~\ref{fig:ntrans}. The direction of neutron transfer from $^{96}$Zr to $^{40}$Ca is expected from the equilibration of the N/Z ratio. On the other hand, the direction of neutron transfer from $^{96}$Zr to $^{48}$Ca is either the result of mass equilibration or the difference in chemical potentials, which is -7.0 MeV for $^{48}$Ca and  -6.2 MeV for $^{96}$Zr for TDSLDA.

It is well established that pairing correlations tend to weaken shell effects, thereby promoting more spherical nuclear shapes.
At the same time, it makes large-amplitude nuclear motion more adiabatic, helping to restore sphericity of the Fermi-surface and suppress collective energy dissipation, enhancing dynamic deformability~\cite{Hill:1953,Bertsch:1980,Barranco:1990,Bertsch:1997,Bulgac:2019c}. This means, even if a nucleus's static deformations are the same with and without pairing, as is the case here for $^{96}$Zr, the character of the dynamic deformations can be very different and will affect observables. To re-emphasize the point, deformations and pairing are very connected, and significant caution should be taken when drawing conclusions from studies that neglect pairing. A more stark demonstration of this fact was shown in~\cite{Ebata:2015}, where they showed that "longer" lived rotation states of $^{22}$O + $^{22}$O and $^{52}$Ca + $^{52}$Ca collisions vanish when pairing is included. Note, the model they used was TDHF+BCS/CbTDHFB. 


 Both TDHF and TDSLDA predict a larger fusion threshold energy than is observed experimentally for $^{40}$Ca + $^{96}$Zr collisions by $\approx4$~MeV, see Table~\ref{tbl:ethres}. This is likely a consequence of the specific choice of NEDF, SeaLL1, as predictions using TDHF with the NEDF, SLy6, fall much closer to experiment, within $\approx0.7$~MeV~\cite{Yao:2024}. The exact threshold energy is very sensitive to the tuning of the NEDF parameters, since the tails of the nuclei barely overlap below this energy. As a result, in order to compare our results with the experiment in a consistent manner, we decided to record all observables with respect to the threshold energy. 

\begin{table}[ht]
\centering
\caption{ \label{tbl:ethres} $E_{\mathrm{Threshold}}$ represents the highest energy before fusion occurs. Experimental results are taken from~\cite{scamps2018b} using data of Ref.~\cite{Timmers:1998}. }
\begin{tabular}{c c c c}
\hline \hline
Reaction & Type & $E_{\mathrm{Threshold}}$ [MeV]& Exp.  \\ \hline \hline
 $^{40}\text{Ca} + {}^{96}\text{Zr}$ & TDHF & $98.4 \pm 0.5$ & 94.3 \\ 
 $^{40}\text{Ca} + {}^{96}\text{Zr}$ & TDSLDA & $98.6 \pm 0.5$ & 94.3 \\ 
  $^{48}\text{Ca} + {}^{96}\text{Zr}$ & TDHF & $95.5 \pm 0.5 $ & -- \\
 $^{48}\text{Ca} + {}^{96}\text{Zr}$ & TDSLDA & $95.5 \pm 0.5$ & --\\ \hline \hline
\end{tabular}
\end{table}

\begin{figure}[h] \includegraphics[width= \columnwidth]{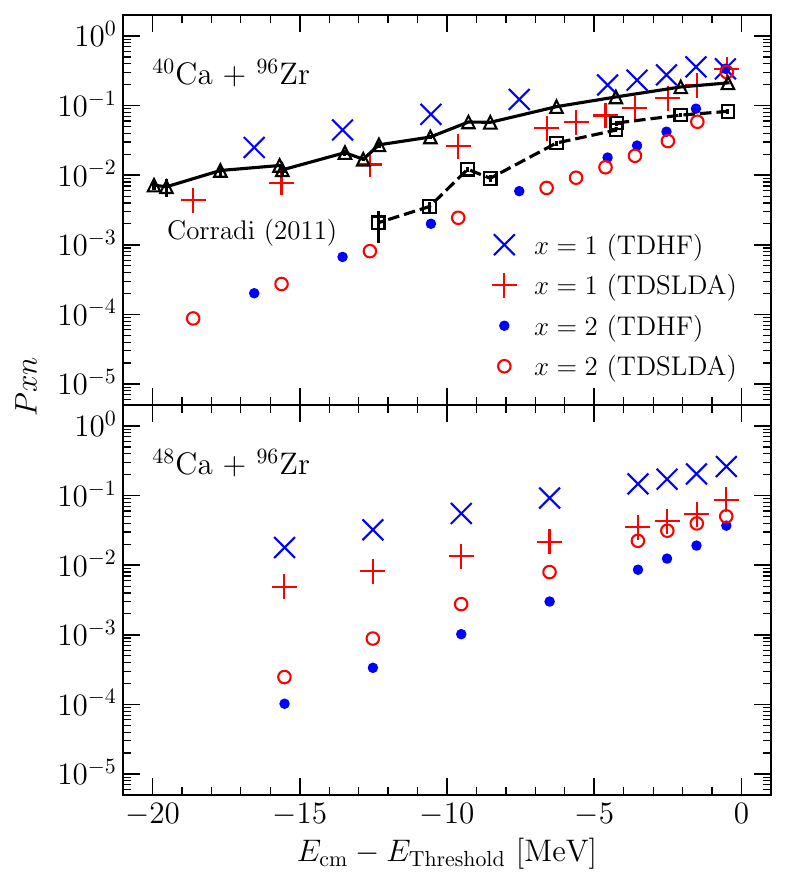}  \caption{ \label{fig:nprob} The one- and two-neutron transfer probabilities are shown as a function of the collision energy. The transfer is calculated after particle projection, and goes from $^{96}$Zr to the Ca isotopes. The black square and triangle markers show results from~\textcite{Corradi:2011} (experiment). The black triangles represent $x=1$, the one particle transfer probability, while the black squares represent $x=2$, the two particle transfer probability. The experimental results are at a finite impact parameter, which we have converted into an effective energy for central collisions, assuming a Rutherford trajectory, to perform the comparison, see also the Supplemental materials~\cite{Supplement:2026}. For backward scattering, this approximation is expected to be reasonable.}  \end{figure}

\begin{figure}[h] 
\includegraphics[width= \columnwidth]{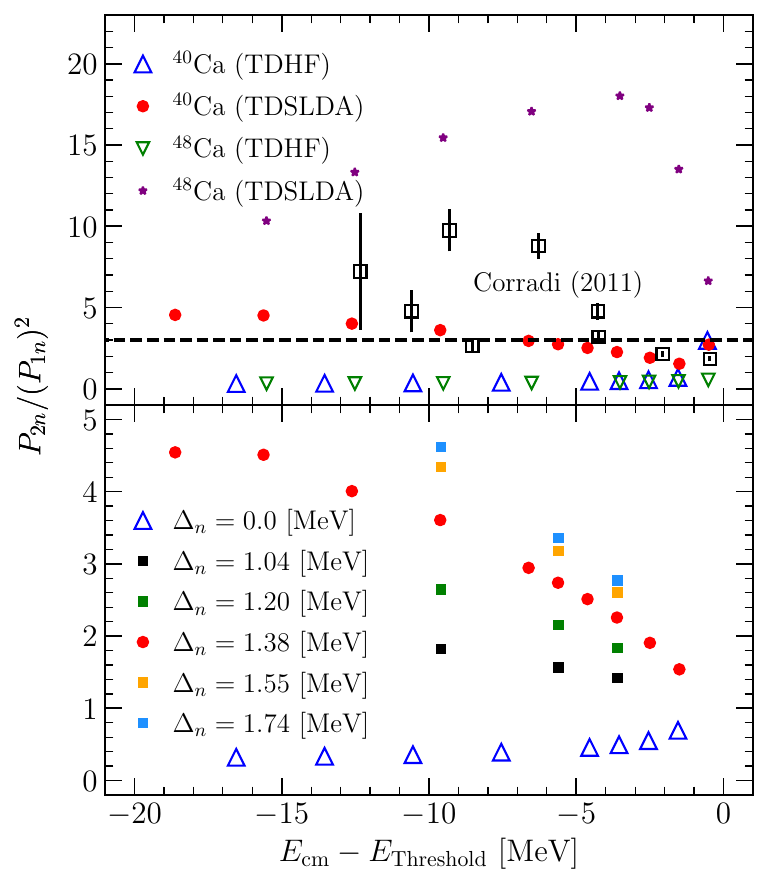} 
\caption{ \label{fig:pratio} 
(Top panel) The enhancement ratio $P_{2n} / P_{1n}^2$ is shown as a function of the collision energy. The reaction type, $^{48,40}$Ca + $^{96}$Zr, is labeled by the Ca isotope. The black squares show results from~\textcite{Corradi:2011} (experiment). The experimental error is estimated by treating $P_{2n},P_{1n}$ as independent observables. The black line represents the value $P_{2n} / P_{1n}^2 = 3$\footnote{
For $^{40}$Ca, at the point closest to the threshold energy, the enhancement ratio increases for both TDSLDA and TDHF. 
This results from the average transfer probability exceeding one, where the concept of an ``enhancement'' ratio $P_{2n}/P_{1n}^2$ ceases to be a probe of correlation. This energy is also the most sensitive to deformations, and hence is not the ideal energy for probing pair correlations.
}.
(Bottom panel) The enhancement ratio is shown for $^{40}$Ca+$^{96}$Zr for different pairing strengths, characterized by the initial pairing gap in $^{96}$Zr. The default value for TDSLDA is given by the red circles. }  \end{figure}

\begin{figure}[h]
    \centering
    \includegraphics[width=\columnwidth]{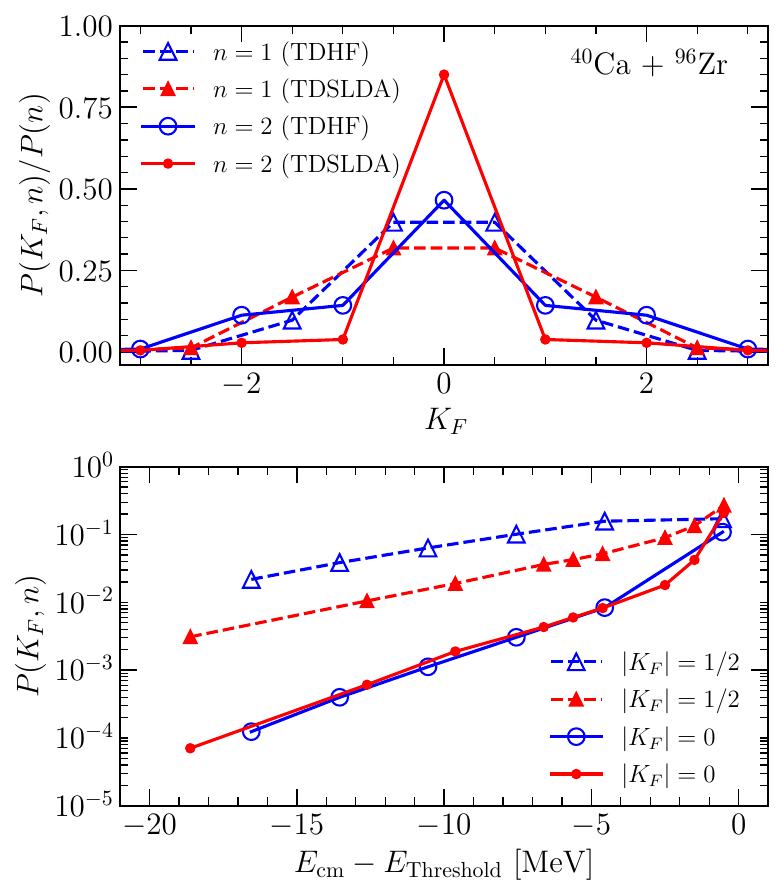}
    \caption{
        \label{fig:pspins}
(Top panel) $K$-distribution in one of the fragments for each transfer channel with and without pairing at a center-of-mass energy $E_{\rm c.m.} = 94$~MeV. In the case of no transfer, n = 0, both nuclei remain exclusively in the $K_F = 0$ channel. (Bottom panel) probability of $|K_F| = 1/2$ and $|K_F| = 0$ spin modes are shown as a function of the collision energy for the 1 and 2 neutron transfer channels.  }
\end{figure}

To compute transfer probabilities microscopically, a particle-number projection method was proposed, within the framework of time-dependent Hartree-Fock (TDHF) theory, by~\textcite{Simenel:2010}. After the reaction, a projection operator is applied to extract the probability that the system is in a state with a specific number of nucleons within a spatial region associated with one of the outgoing fragments. This approach has been extensively applied to multinucleon transfer reactions, see Ref.~\cite{sekizawa2019} and references therein. The projection method has been generalized to include pairing correlations via the BCS approximation, particularly when one of the fragments is superfluid~\cite{Scamps:2013}.  We only project on the neutron sector.

Using this method, we extracted the one- and two-neutron transfer probabilities for $^{40,48}$Ca + $^{96}$Zr collisions, see Fig.~\ref{fig:nprob}. For all energies, the one-neutron transfer probabilities are smaller when pairing correlations are included, consistent with the trend for the mean transfer of neutrons, which is controlled by the dynamic deformability of $^{96}$Zr described above.  In comparison, the probabilities of two-neutron transfer are greater for the $^{48}$Ca + $^{96}$Zr reactions (shown in the bottom panel) or roughly equal for the $^{40}$Ca + $^{96}$Zr reactions (shown in the top panel) when pairing correlations are included.  In all cases, the ratio $P_{2n} / P_{1n}^2$ is enhanced by pairing correlations, shown in Fig.~\ref{fig:pratio}. For $^{40}$Ca + $^{96}$Zr, the enhancement ratio for TDSLDA shows good agreement with experiment.  In previous studies using the BCS approximation, the ratio was underestimated~\cite{Zhang:2026,Scamps:2015b}. This discrepancy is suspected to arise from the limitations of the BCS approximation in accurately capturing pairing correlations. Interestingly, we predict a much larger enhancement ratio $^{48}$Ca + $^{96}$Zr than $^{40}$Ca + $^{96}$Zr. This can be understood by examining the final neutron pairing gaps of the system, which are larger in the $^{48}$Ca + $^{96}$Zr reactions. At $E_{cm}$ = 95 MeV, for $^{48}$Ca + $^{96}$Zr reactions, the final pairing gaps are given by $\Delta_n$ = 0.22 MeV for $^{48}$Ca and $\Delta_n$ = 1.13 for $^{96}$Zr. At the same energy, for $^{40}$Ca + $^{96}$Zr reactions, the final pairing gaps are given by $\Delta_n$ = 0.07 MeV for $^{40}$Ca and $\Delta_n$ = 1.02 for $^{96}$Zr. Note, the initial pairing gaps are zero for both Calcium isotopes. 

The enhancement ratio might be one of the clearest and most sensitive experimental observables for constraining both the strength and the treatment of pairing correlations in theory. To quantify this effect, we modified the bare pairing coupling constant for both neutrons and protons. We see the enhancement ratio is strongly affected by the strength of pairing, which is characterized by the pairing gap
\begin{equation}
\Delta_n \equiv \frac{\int d^3 r |\Delta_n(\bm{r}) |n_n(\bm{r})}{\int d^3 r n_n(\bm{r})}, 
\end{equation}
as shown in Fig.~\ref{fig:pratio}. We have also checked and confirmed that it is indeed the most sensitive observable to pairing, with further details provided in the Supplemental materials~\cite{Supplement:2026}. 

Another important question, related to pair transfer, is which states are populated. Is the transfer directly to the ground state of the fragments? See for instance~\cite{Montanari:2016}.
In order to more precisely characterize the one- and two-neutron transfer processes, we employ an additional projector on the angular momentum $K$. This projector is defined in~\cite{Ring:2004}. In our case, we are interested in the $K$-distribution of neutrons in each of the transfer channels. Therefore, we combine particle number and angular momentum projection.

The $K$-distribution in each neutron transfer channel is shown in Fig.~\ref{fig:pspins} and provides insight into the pair-breaking mechanism during the transfer reaction. In fully paired even-even systems, and in the absence of triaxial deformation, each nucleon with a given $K$ is paired with a nucleon of $-K$, resulting in a system with total $K = 0$. In the no-transfer channel, the system remains fully paired, and the probability of pair breaking is of the order of $10^{-4}$. This is also the case in the no-pairing calculation, where the last shells of neutrons and protons for both fragments are fully occupied. 
The one-neutron transfer channel necessarily involves the breaking of a pair. Interestingly, the transfer predominantly occurs from or to states with $|K| = 1/2$, which correspond to wave functions that extend along the $z$-axis—the axis of elongation during the collision—and are therefore more likely to participate in the transfer. In the case of pair transfer, the final state is dominated by the $K$~=~0 channel. To a lesser extent, this is also the case without pairing because the $|K|$~=~1/2 sequential transfer is favored. This trend holds at all energies, except at the threshold energy, as shown in Fig.~\ref{fig:pspins}.

The $K$-distribution is similar to the one found in fission \cite{Scamps:2023}, with a dominant peak at $|K|=0$. However, the transfer under the barrier in collisions involves less excitation energy in the fragments relative to fission, and so the pair-breaking mechanism is less important. For $^{40}$Ca + $^{96}$Zr at $E_{\mathrm{cm}} = 94$ MeV, the total excitation energy is $\approx 1$ MeV ($\approx$10 keV per nucleon), while for the fission of $^{240}$Pu, within the same framework, it is $\approx$ 30-40 MeV ($\approx$100 keV per nucleon)~\cite{Bulgac:2019c}. 

In summary, we have performed the first simulations of $^{40,48}$Ca + $^{96}$Zr central collisions at energies below the fusion threshold, using TDSLDA. We demonstrated that pairing correlations have a strong impact on the dynamic deformability of nuclei, which in turn affects observables, such as the average neutron transfer. We also performed the first microscopic extraction of the enhancement ratio, which is in agreement with the experiment. This ratio is one of the best observables for constraining pairing in theoretical models. This enhancement is explained by the two-neutron transfer predominantly occurring in the $K=0$ channel, which corresponds to correlated pair transfer.

\begin{acknowledgments}
I.A. thanks Aurel Bulgac for his feedback regarding the manuscript.
I.A. was partially supported by the U.S. Department of Energy through the Los Alamos National Laboratory.
The Los Alamos National Laboratory is operated by Triad National Security, LLC, for the National Nuclear Security Administration of the U.S. Department of Energy Contract No. 89233218CNA000001.
This work was in part supported by Michigan State University through the Reaction Theory Initiative.
K.G. was support in part by the U.S. Department of Energy under Award Numbers DE-SC0013365 (Office of Science, Office of Nuclear Physics) and DE-SC0023175 (Office of Science, NUCLEI SciDAC-5 collaboration).
A.M. and P.M. were supported by the Polish National Science Center under Grant No. UMO-2021/43/B/ST2/01191.
A.M was supported by the Mobility IX action at his affiliation.
We acknowledge the Polish highperformance computing infrastructure, PLGrid, for
awarding this project access to the LUMI supercomputer, owned by the EuroHPC Joint Undertaking, hosted
by CSC (Finland), and the LUMI consortium through PLL/2023/05/016762.
I.A. also gratefully acknowledges partial support provided by the Advanced Simulation and Computing (ASC) Program. We are also thankful to the HPC resources of IDRIS under Allocation No. 2024-AD010515531R1 made by GENCI to perform the angular momentum projection. I.A. and A.M. implemented the program, performed the TDDFT calculations, and processed the data. G.S. performed the angular momentum projections. All authors participated in the discussion of the results and writing of the manuscript. 
\end{acknowledgments}

\clearpage

\providecommand{\selectlanguage}[1]{}
\renewcommand{\selectlanguage}[1]{}

\bibliographystyle{apsrev4-1}  
\bibliography{CollisionsBib}

\end{document}


\title{Supplemental Materials}
  
\author{Ibrahim Abdurrahman}
  \affiliation{ Facility for Rare Isotope Beams, Michigan State University, East Lansing, Michigan 48824, USA}
 \affiliation{ Theoretical Division, Los Alamos National Laboratory, Los Alamos, New Mexico 87545, USA}

 \author{Andrzej Makowski}
 \affiliation{Faculty of Physics, Warsaw University of Technology, \\
ulica Koszykowa 75, Warsaw, 00-662, Poland}

 \author{Guillaume Scamps}
 \affiliation{ Université de Toulouse, CNRS/IN2P3, L2IT, Toulouse, France }

 \author{Kyle Godbey}
 \affiliation{ Facility for Rare Isotope Beams, Michigan State University, East Lansing, Michigan 48824, USA}

  \author{Piotr Magierski}
 \affiliation{Faculty of Physics, Warsaw University of Technology, \\
ulica Koszykowa 75, Warsaw, 00-662, Poland}
 \affiliation{Physics Department, University of Washington, \\ 3910 15th Ave. NE, Seattle, WA 98195-1560, USA}
\date{\today}

\preprint{LA-UR-25-31452}

\maketitle


 \vspace{0.5cm}

\section*{Density Profiles}

\begin{figure}[h] 
\includegraphics[width=0.85\columnwidth]{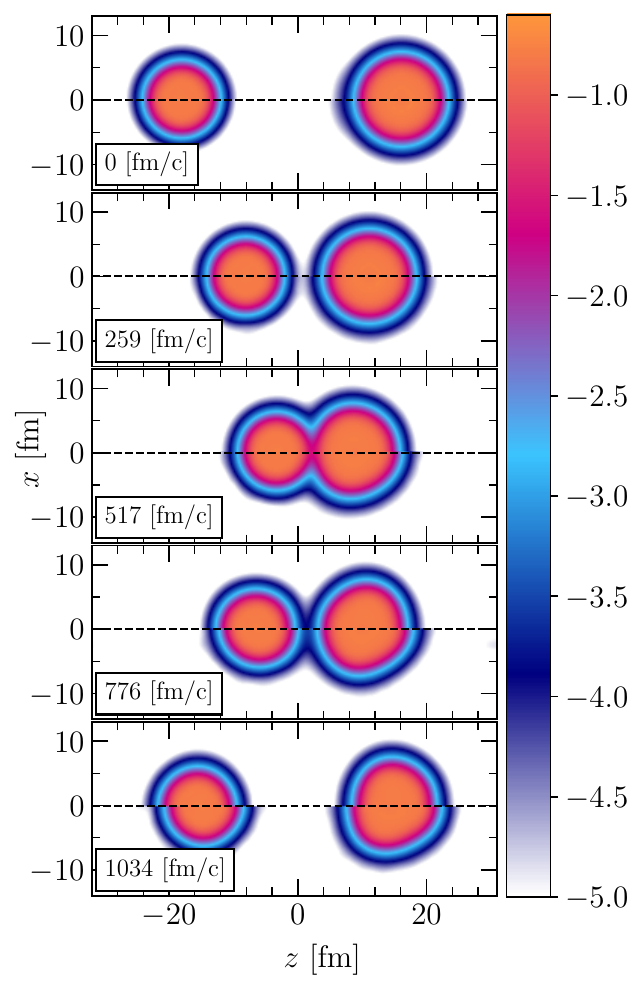}  
\caption{ 
\label{fig:pdensities}  Slices of the total number density in log scale are shown for $^{48}$Ca + $^{96}$Zr collisions at $E_\mathrm{cm} = 95.0 $ MeV for different times. The top half of each panel are TDSLDA simulations while the bottom half each panel are TDHF simulations. The nuclei interact slightly longer for TDHF. 
}  
\end{figure}

All simulations were performed on a $28^2 \times 64$ Cartesian lattice with a 1 fm spacing using a spherical cutoff with cutoff energy $E_{\mathrm{cutoff}} = 102$ MeV. All collisions started with the nuclei separated by 34 fm, with the exception of the TDSLDA runs of $^{40}$Ca + $^{96}$Zr, which started at a separation distance of 36 fm. Fig.~\ref{fig:pdensities} shows the slices of the total nuclear density of $^{48}$Ca + $^{96}$Zr collisions, with and without pairing correlations. After the collision, it is clear the character of the deformation of $^{96}$Zr is different for TDHF versus TDSLDA.  

\section*{Effective head-on collision}

In Ref.~\textcite{Corradi:2011}, the experiment is conducted at a fixed scattering angle of \(\theta_{\rm c.m.} = 140^\circ\). This corresponds to a classical Rutherford trajectory with a finite impact parameter. Following Ref.~\cite{Scamps:2013}, we assume that the transfer probability depends only on the distance of closest approach \(D\), which is given by:

\begin{align}
D = \frac{Z_P Z_T e^2}{2 E_{\rm c.m.}} \left(1 + \frac{1}{\sin(\theta_{\rm c.m.}/2)}\right).
\end{align}

All calculations in this work are presented as a function of an effective center-of-mass energy for head-on collisions:

\begin{align}
E_{\rm c.m.}' = \frac{E_{\rm c.m.}}{1 + 1/\sin(\theta_{\rm c.m.}/2)}.
\end{align}

\section*{Results for Unshifted Energy}

\begin{figure}[H] \includegraphics[width=0.85\columnwidth]{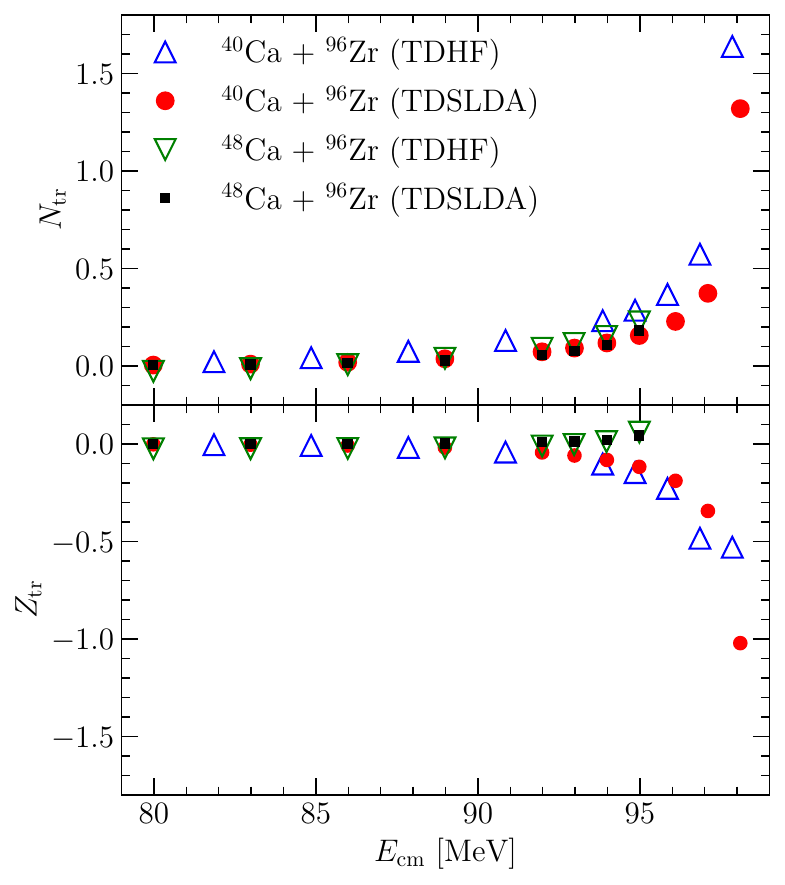}  \caption{ \label{fig:ntrans_sup}  (Top panel) The mean neutron transfer to the Ca isotopes is shown as a function of the collision energy. (Bottom panel) same for protons. }  \end{figure}

\begin{figure}[H] \includegraphics[width=0.85\columnwidth]{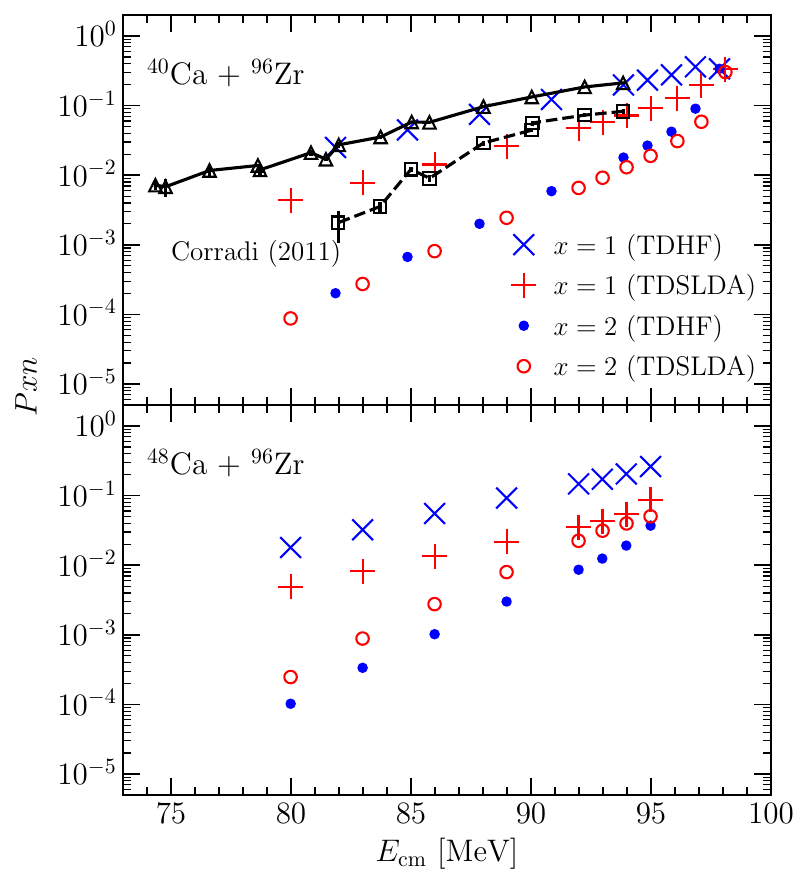}  \caption{ \label{fig:nprob_sup}  The one- and two-neutron transfer probabilities are shown as a function of the collision energy. The black square and triangle markers show results from~\textcite{Corradi:2011} (experiment). The black triangles represent $x=1$, the one particle transfer probability, while the black squares represent $x=2$, the two particle transfer probability.}  \end{figure}

\begin{figure}[H] \includegraphics[width=0.85\columnwidth]{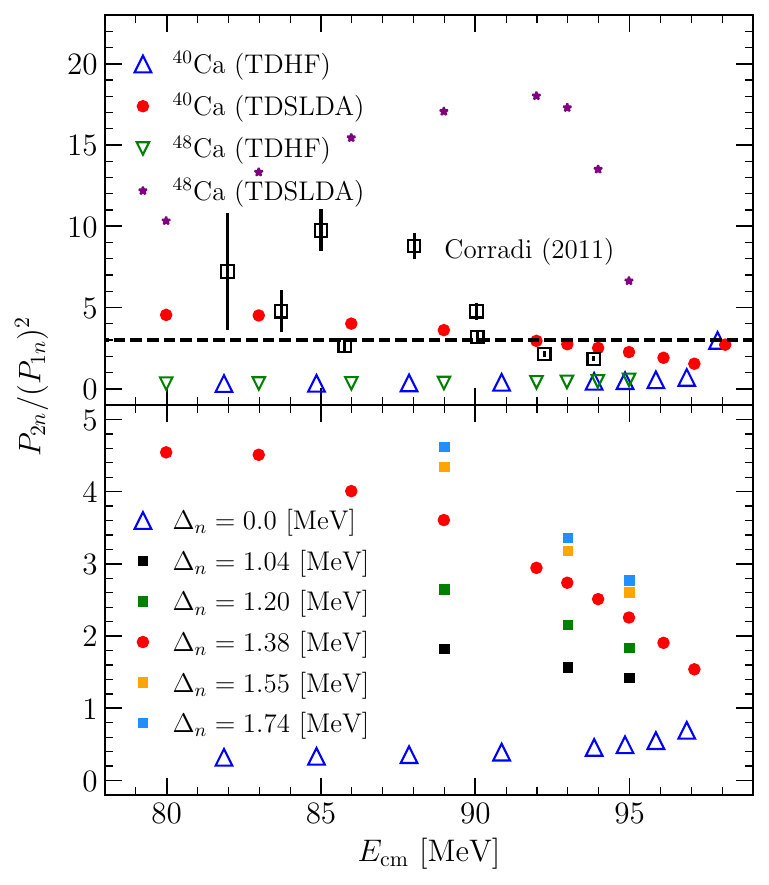}  \caption{ \label{fig:pratio_sup} (Top panel) The enhancement ratio $P_{2n} / P_{1n}^2$ is shown as a function of the collision energy. (Bottom panel) the enhancement ratio is shown for $^{40}$Ca+$^{96}$Zr for different pairing strengths, characterized by the initial pairing gap in $^{96}$Zr.}  \end{figure}

\begin{figure}[h]
    \centering
    \includegraphics[width=\columnwidth]{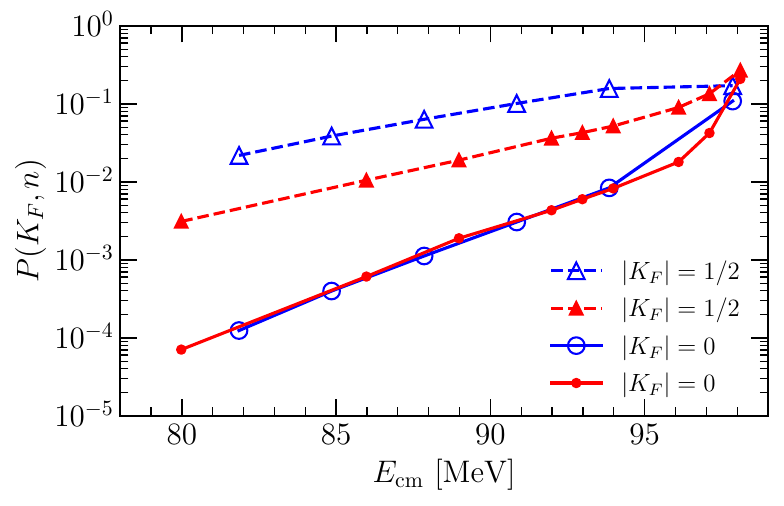}
    \caption{
        \label{fig:pspins_sup}
Probability of $|K_F| = 1/2$ and $|K_F| = 0$ spin modes are shown as a function of the collision energy for the 1 and 2 neutron transfer channels, for the $^{40}$Ca + $^{96}$Zr reaction. The one transfer channel is given by the triangles and the two transfer channel is given by the circles. The simulations without pairing (TDHF) are given by the blue open markers, and simulations with pairing (TDSLDA) are given by the red closed markers.}
\end{figure}

\begin{figure}[H] \includegraphics[width=0.85\columnwidth]{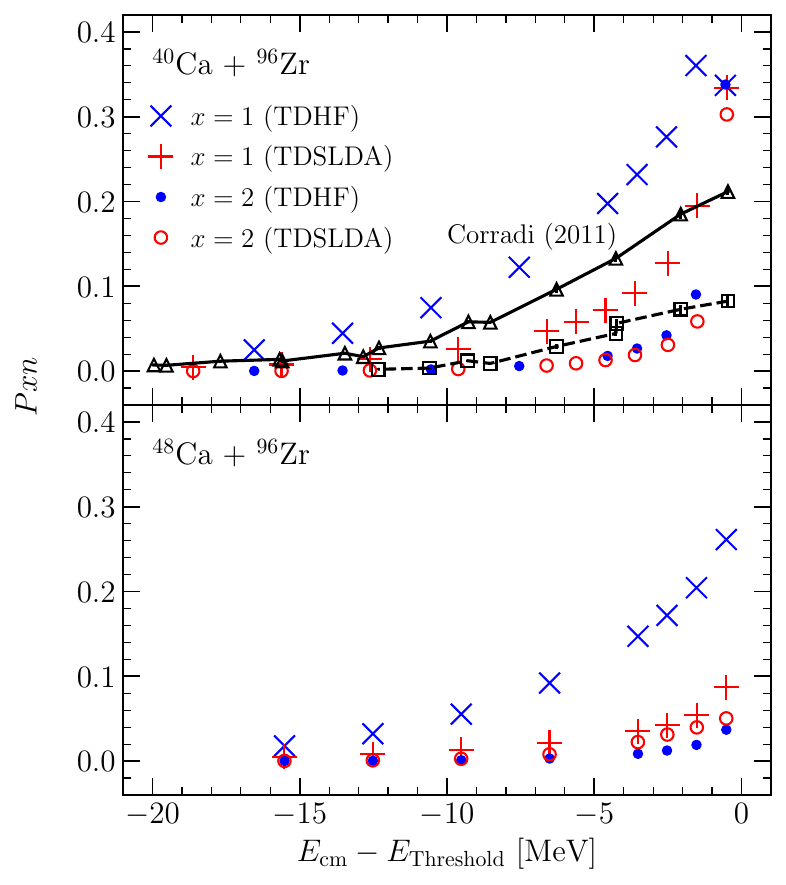}  \caption{ \label{fig:nprob_lin}  The one- and two-neutron transfer probabilities are shown as a function of the collision energy in linear scale with respect to the threshold
energy. The black square and triangle markers show results from~\textcite{Corradi:2011} (experiment). The black triangles represent $x=1$, the one particle transfer probability, while the black squares represent $x=2$, the two particle transfer probability. }  \end{figure}

\begin{figure}[H] \includegraphics[width=0.85\columnwidth]{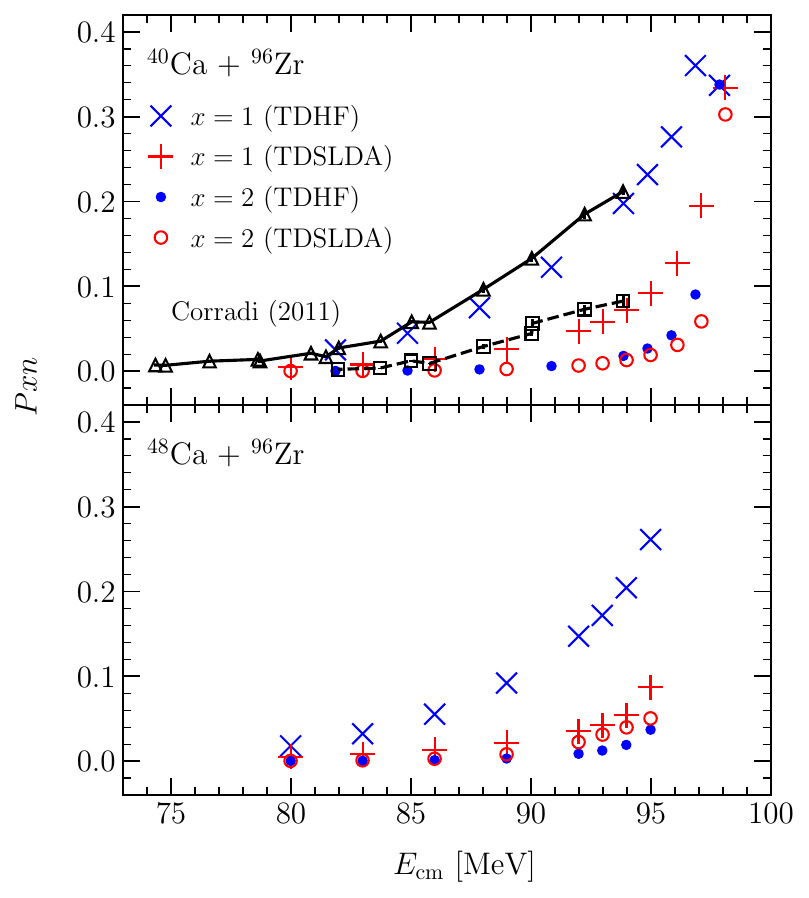}  \caption{ \label{fig:nprob_lin_sup}  The one- and two-neutron transfer probabilities are shown as a function of the collision energy in linear scale. The black square and triangle markers show results from~\textcite{Corradi:2011} (experiment). The black triangles represent $x=1$, the one particle transfer probability, while the black squares represent $x=2$, the two particle transfer probability.}   \end{figure}

For convenience, we have included Figures~\ref{fig:ntrans_sup},~\ref{fig:nprob_sup},~\ref{fig:pratio_sup}, and~\ref{fig:pspins_sup}, which correspond to similar figures in the main manuscript, except the energy is no longer shifted by the threshold energy. Figures~\ref{fig:nprob_lin} and~\ref{fig:nprob_lin_sup} show the one and two particle transfer probabilities in linear scale, with and without the energy shift.

\section*{Results for Pairing Strength}

\begin{figure}[H] \includegraphics[width=0.85\columnwidth]{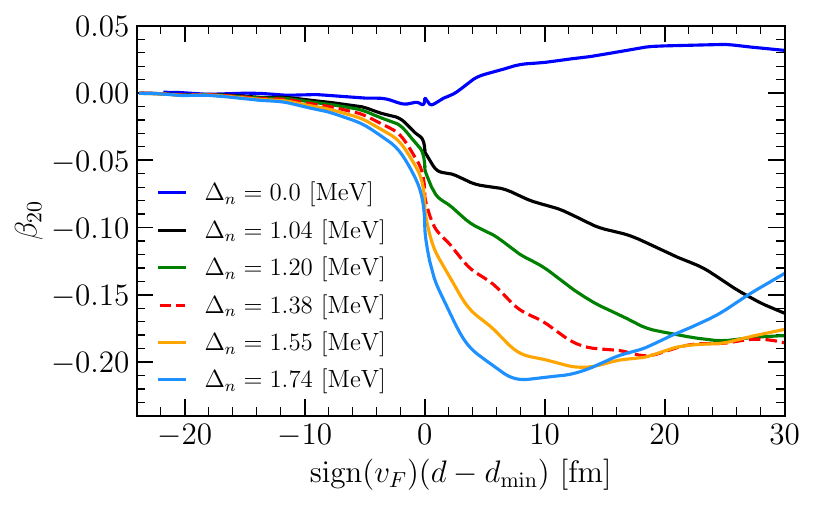}  \caption{ \label{fig:deformpair}  The deformation parameter $\beta_{20}$ of $^{96}$Zr is shown as a function of the separation distance $d = \bm{r}_{cm,R} - \bm{r}_{cm,L}$ between the two nuclei minus the minimum separation distance reached during the collision $d_{\mathrm{min}}$ for different pairing strengths, characterized by the initial pairing gap in $^{96}$Zr. The reaction is $^{40}$Ca + $^{96}$Zr at $E_{cm}$ = 95 MeV.}  \end{figure}

\begin{figure}[H] \includegraphics[width=0.85\columnwidth]{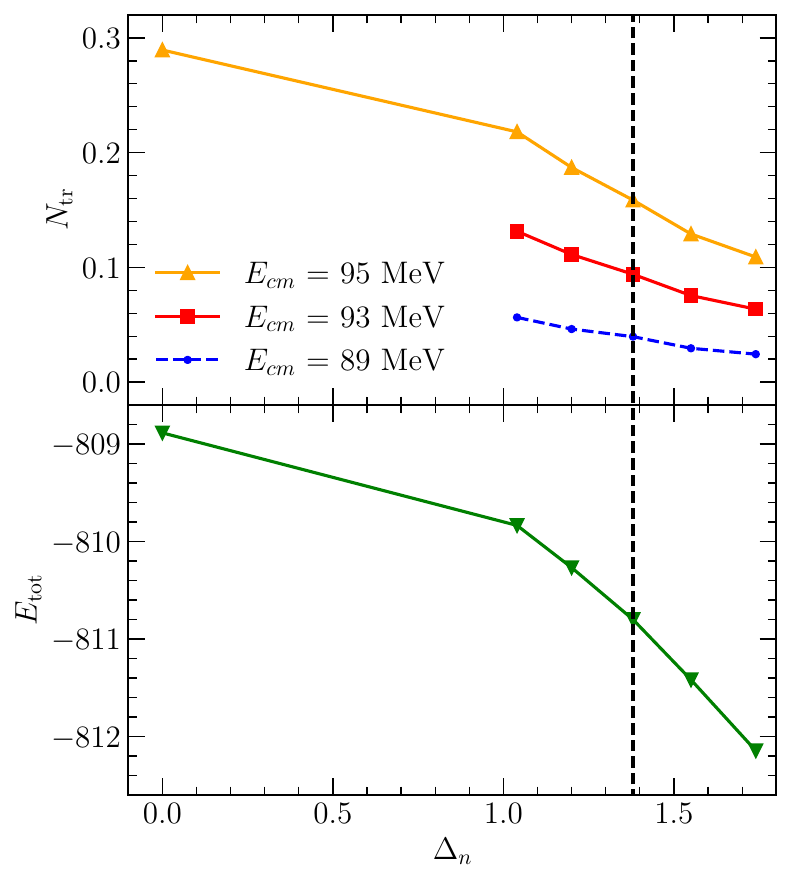}  \caption{ \label{fig:transpair}  (Top panel) the mean transfer of neutrons to the Ca isotopes is shown as a function of the initial pairing gap in $^{96}$Zr, corresponding to different pairing strengths. (Bottom panel) the initial binding energies of $^{96}$Zr are similarly shown. The black dashed line corresponds to the standard value of the pairing strength. The reaction is $^{40}$Ca + $^{96}$Zr. }  \end{figure}

Figures~\ref{fig:deformpair} and~\ref{fig:transpair} show how observables vary as a function of the pairing strengths, characterized by the initial pairing gap in $^{96}$Zr. The bare pairing coupling constant for both neutrons and protons were chosen at $g_0 = 0, 180, 190, 200, 210,$ and $220$, with $g_0 = 200$ being the default value for the EDF SeaLL1 with a spherical cutoff. These correspond to initial pairing gaps in $^{96}$Zr of $\Delta_n = 0.0, 1.04, 1.20, 1.38, 1.55,$ and $1.74$ MeV. SeaLL1 uses strictly volume pairing. 

The deformation is defined by,
\begin{equation}
    \beta_{20} =  \frac{4 \pi}{3 R_0^2 A^{5/3}} \int  r'^2 Y_{20}(x,y,z')\rho(\bm{r}') d^3 r ,
\end{equation}
where $\rho(\bm{r}')$ and $A$ represent the number density and mass of the Zr-like fragment, with $r'^2 = x^2 + y^2 + z'^2$, $z'  = z - z_{\mathrm{cm}}$, and $z_{\mathrm{cm}}$ representing the center of mass of the Zr-like fragment. The collision axis is along z. $R_0 = 1.2 $ The spherical harmonic is defined by,
\begin{equation}
Y_{20}(x,y,z') = \sqrt{\frac{5}{16\pi}} \Big(\frac{3 z'^2 - r'^2}{r'^2}\Big).
\end{equation}

\section*{Details on Angular Momentum Projection}

This projector on the z-component of the spin of the fragments is defined as~\cite{Ring:2004,ZHANG:2025}:
\begin{align}
\hat{P}_{K_F}(K_F) = \frac{1}{4\pi} \int_0^{4\pi} e^{-i \varphi (\hat{j}_{zF} - K_F)} \, d\varphi,
\end{align}
where the operator $e^{-i \varphi \hat{j}_{zF}}$ rotates the fragment by an angle $\varphi$ around the $z$-axis, which is the axis of deformation of the fragments during the collision. $F$ design the fragment $L$ (left) or $R$ (right).
Since the system initially possesses axial symmetry, the projection $j_z$ of the total angular momentum of the full system on the $z$-axis is exactly zero.
Consequently, it is not necessary to perform a projection on the total $K$ of the system, and the projection operators satisfy the relation $P_{K_L}(K) = P_{K_R}(-K) = P_{K_L}(-K)$, where $K_L$ and $K_R$ denote the $K$-quantum numbers of the left and right fragments, respectively. 
Note that the present interpretation of $j_{zF}$ as the $K$-quantum number is only valid in the case of a head-on collision.

In our case, we are interested in the $K$-distribution of neutrons in each of the transfer channels. We therefore compute the conditional probability on the many-body wave function,
\begin{align}
P(K_F,N_F) = \frac{\langle \Psi | \hat{P}_{K_F}(K_F) \hat{P}_{N_F}(N_F) \hat{P}_{N_0}(N_0) | \Psi \rangle}{\langle \Psi | \hat{P}_{N_0}(N_0) | \Psi \rangle}.
\end{align}
With $\hat{P}_{N_F}(N_F)$ the projector on the number of neutrons in the fragment $F$ and $\hat{P}_{N_0}(N_0)$ the projector on the total number of neutrons in the system.
Note that the projection method itself is exact. As a result, for states where $N_F$ is even (odd), the probability of finding a half-integer (integer) value of $K_F$ is exactly zero. 

The projection for angular momentum and particle numbers are done using the Pfaffian formulae of Ref.~\cite{Scamps:2013,Bertsch12} which have been proven equivalent to other formulae used in the literature~\cite{Robledo:2025}.

\section*{Comparison with TDHF+BCS Results}

In Ref.~\cite{Scamps:2013}, the TDHF+BCS+FOA framework is employed, whereas our work uses the TDHFB/TDSLDA  approach. The BCS approximation assumes that the interaction is diagonal in the canonical basis, which simplifies the solution of the static and dynamic HFB/TDHFB equations but introduces several limitations:
\begin{enumerate}
\item Pairing Gap Adjustment: The effective pairing interaction must be artificially increased to reproduce the experimental pairing gap (e.g., 3 or 5 points).
\item Gas Problem: The BCS approximation leads to the "gas problem," necessitating a strict energy cutoff at zero energy.
\item Violation of Continuity: The continuity equation is not preserved in the TDHF+BCS theory.
Additionally, Ref.~\cite{Scamps:2013} uses the Frozen Occupation Approximation (FOA), which means dynamical pairing effects are not included in the calculations. Our approach avoids these limitations by using TDHFB/TDSLDA, which provides a more consistent and physically accurate treatment of pairing dynamics.
\end{enumerate}

\section*{Minimum Distances}

The minimum distances $d_\mathrm{min}$ are shown for all reactions and energies in Table~\ref{tbl:dmin}.

\begin{table}[ht]
\centering
\caption{ \label{tbl:dmin} Across the columns, the reaction, framework, collision energy in the center of mass frame, minimum distance, and estimated minimum distance, are shown. }
\begin{tabular}{c c c c c}
\hline \hline
Reaction & Type & $E_{cm}$ & $d_\mathrm{min}$ & $e^2 \frac{Z_L Z_R}{E_{cm}}$  \\ \hline \hline
$^{40}\text{Ca} + {}^{96}\text{Zr}$ & TDHF & 81.85 & 14.25 & 14.07 \\
$^{40}\text{Ca} + {}^{96}\text{Zr}$ & TDHF & 84.85 & 13.77 & 13.58 \\
$^{40}\text{Ca} + {}^{96}\text{Zr}$ & TDHF & 87.85 & 13.33 & 13.11 \\
$^{40}\text{Ca} + {}^{96}\text{Zr}$ & TDHF & 90.85 & 12.92 & 12.68 \\
$^{40}\text{Ca} + {}^{96}\text{Zr}$ & TDHF & 93.85 & 12.51 & 12.27 \\
$^{40}\text{Ca} + {}^{96}\text{Zr}$ & TDHF & 94.85 & 12.37 & 12.14 \\
$^{40}\text{Ca} + {}^{96}\text{Zr}$ & TDHF & 95.85 & 12.23 & 12.02 \\
$^{40}\text{Ca} + {}^{96}\text{Zr}$ & TDHF & 96.85 & 12.08 & 11.89 \\
$^{40}\text{Ca} + {}^{96}\text{Zr}$ & TDHF & 97.85 & 11.92 & 11.77 \\ \hline
$^{40}\text{Ca} + {}^{96}\text{Zr}$ & TDSLDA & 79.98 & 14.56 & 14.40 \\
$^{40}\text{Ca} + {}^{96}\text{Zr}$ & TDSLDA & 82.98 & 14.07 & 13.88 \\ 
$^{40}\text{Ca} + {}^{96}\text{Zr}$ & TDSLDA & 85.98 & 13.61 & 13.40 \\
$^{40}\text{Ca} + {}^{96}\text{Zr}$ & TDSLDA & 88.98 & 13.18 & 12.95 \\
$^{40}\text{Ca} + {}^{96}\text{Zr}$ & TDSLDA & 91.98 & 12.77 & 12.52 \\
$^{40}\text{Ca} + {}^{96}\text{Zr}$ & TDSLDA & 92.98 & 12.64 & 12.39 \\
$^{40}\text{Ca} + {}^{96}\text{Zr}$ & TDSLDA & 93.98 & 12.50 & 12.26 \\
$^{40}\text{Ca} + {}^{96}\text{Zr}$ & TDSLDA & 94.98 & 12.37 & 12.13 \\
$^{40}\text{Ca} + {}^{96}\text{Zr}$ & TDSLDA & 96.10 & 12.22 & 11.99 \\
$^{40}\text{Ca} + {}^{96}\text{Zr}$ & TDSLDA & 97.10 & 12.08 & 11.86 \\
$^{40}\text{Ca} + {}^{96}\text{Zr}$ & TDSLDA & 98.10 & 11.93 & 11.74 \\ \hline
$^{48}\text{Ca} + {}^{96}\text{Zr}$ & TDHF & 79.98 & 14.49 & 14.40 \\
$^{48}\text{Ca} + {}^{96}\text{Zr}$ & TDHF & 82.98 & 13.97 & 13.88 \\
$^{48}\text{Ca} + {}^{96}\text{Zr}$ & TDHF & 85.98 & 13.51 & 13.40 \\
$^{48}\text{Ca} + {}^{96}\text{Zr}$ & TDHF & 88.98 & 13.07 & 12.95 \\
$^{48}\text{Ca} + {}^{96}\text{Zr}$ & TDHF & 91.98 & 12.63 & 12.52 \\
$^{48}\text{Ca} + {}^{96}\text{Zr}$ & TDHF & 92.97 & 12.49 & 12.39 \\
$^{48}\text{Ca} + {}^{96}\text{Zr}$ & TDHF & 93.97 & 12.33 & 12.26 \\
$^{48}\text{Ca} + {}^{96}\text{Zr}$ & TDHF & 94.98 & 12.18 & 12.13 \\ \hline
$^{48}\text{Ca} + {}^{96}\text{Zr}$ & TDSLDA & 79.98 & 14.45 & 14.40 \\
$^{48}\text{Ca} + {}^{96}\text{Zr}$ & TDSLDA & 82.98 & 13.94 & 13.88 \\
$^{48}\text{Ca} + {}^{96}\text{Zr}$ & TDSLDA & 85.98 & 13.48 & 13.40 \\
$^{48}\text{Ca} + {}^{96}\text{Zr}$ & TDSLDA & 88.98 & 13.04 & 12.95 \\
$^{48}\text{Ca} + {}^{96}\text{Zr}$ & TDSLDA & 91.98 & 12.60 & 12.52 \\
$^{48}\text{Ca} + {}^{96}\text{Zr}$ & TDSLDA & 92.98 & 12.46 & 12.39 \\
$^{48}\text{Ca} + {}^{96}\text{Zr}$ & TDSLDA & 93.98 & 12.31 & 12.26 \\
$^{48}\text{Ca} + {}^{96}\text{Zr}$ & TDSLDA & 94.98 & 12.16 & 12.13\\ \hline \hline
\end{tabular}
\end{table}

\section*{PES of $^{96}\text{Zr}$}

\begin{figure}[H] \includegraphics[width=0.85\columnwidth]{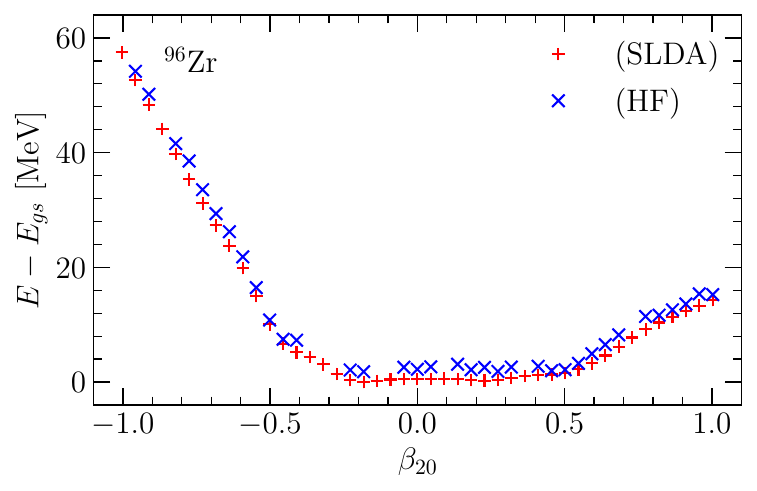}  \caption{ The energy of $^{96}$Zr is shown as a function of $\beta_{20}$ with respect to it's ground state energy. The red plus markers include pairing, while the blue cross markers do not including pairing. Both SLDA and HF energies are taken with respect to the energy of the SLDA ground state solution. } \label{fig:pes}   \end{figure}

In Fig.~\ref{fig:pes} the 1-D potential energy surface (PES) is shown for $^{96}$Zr with respect to it's quadrupole moment for the SeaLL1 EDF. The surface is primarily flat between $\beta_{20} = -0.5$ to  $\beta_{20} = 0.5$ allowing for the nucleus to deform dynamically. 

\providecommand{\selectlanguage}[1]{}
\renewcommand{\selectlanguage}[1]{}

\bibliographystyle{apsrev4-1}  
\bibliography{CollisionsBib}